\documentclass[twocolumn,showkeys,preprintnumbers,amsmath,amssymb]{revtex4}

\usepackage{natbib}
\usepackage{amsmath,amssymb,amsfonts,amsthm,euscript,changebar,subfigure,epsfig,amscd,stmaryrd}
\usepackage{graphicx}
\usepackage{dcolumn}
\usepackage{bm}
\usepackage[colorlinks]{hyperref}
\usepackage{epstopdf}

\hypersetup{
    colorlinks=true,       
    linkcolor=black,       
    citecolor=black,       
    filecolor=black,       
    urlcolor=black         
}

\newtheorem{Example}{Example}

\begin{document}

\title{On the stability analysis in the transition to turbulence problem}

\author{R. Krechetnikov}
\affiliation{Department of Mathematics \\
University of Alberta, Edmonton, Canada}
\author{J.~E. Marsden}
\affiliation{Control and Dynamical Systems \\
California Institute of Technology, Pasadena, USA}

\date{\today}

\begin{abstract}
In this note, which is of general stability theory interest, we
discuss some of the key assertions usually stated in the context
of the transition to turbulence problem. In particular, the two
main points made here in the setting of the transition problem are
(i) the crucial dependence of the stability results on whether the
problem is considered on infinite or semi-infinite domain, and
(ii) the energy conservation by the nonlinear terms of the
Navier-Stokes equations. As an application, we demonstrate that
the Couette flow, when analyzed in the mathematical setting most
correctly reflecting the way the experiments are usually done, is
spectrally unstable for finite Reynolds numbers in apparent
contradiction to the commonly accepted classical century-old
results. Also, the interrelation of various stability notions, the
effects of infinite dimensionality, the covariant nature of the
transition phenomena and how non-normality of the linear operators
and finite-amplitude instability fit into this picture are
discussed as well.
\end{abstract}



\maketitle

\tableofcontents

\section{Introduction}

When exploring physical phenomena, one naturally uses both
experimental and theoretical tools to gain as complete
understanding as possible. If understanding of the physics is
relatively complete, then not fully rigorous mathematical
description usually suffices. If, however, the phenomenon is as
complicated, as the transition to turbulence problem, where
despite numerous experimental studies there is no firm grasp of
physics, then the progress inevitably relies upon rigorous
mathematical tools. Since, all the equations we use in physics are
phenomenological, then only a constructive symbiosis of
experimental and rigorous theoretical tools allows us to
understand the validity and the range of applicability of these
equations. The Navier-Stokes equations (NSEs) for incompressible
fluid, we discuss here,
\begin{subequations}
\label{NSEs}
\begin{align}
{\partial \mathbf{u} \over \partial t} + \mathbf{u} \, \nabla
\mathbf{u} &= - {1 \over \rho} \nabla p + \nu \Delta
\mathbf{u} \ \mathrm{in} \ \Omega \times (0,\infty) \\
\nabla \cdot \mathbf{u} &= 0 \ \mathrm{in} \ \Omega \times
(0,\infty)
\end{align}
\end{subequations}
are not an exception, since their analytical properties are still
not fully understood. System \eqref{NSEs} is supplied, as usual,
with the appropriate boundary conditions at $\partial \Omega$ and
initial conditions at $t=0$.

The transition to turbulence problem has long been considered in
the realms of hydrodynamic stability theory, the key question of
which can be formulated as follows: what happens to a given fluid
flow (base state) under the influence of disturbances. If the flow
is robust under the influence of all possible disturbances, it is
called (Lyapunov) stable and can be expected to be observed in
Nature. If there are perturbations which start to grow, the flow
is called unstable and thus is expected to break up. If there is
some finite critical amplitude of disturbance, beyond which the
flow is unstable, then it is called finite-amplitude unstable. One
can also enrich this picture by quantifying not only amplitudes of
the critical disturbances, but also their geometry in some
appropriate infinite-dimensional phase space. Despite this
seemingly straightforward view of stability, there is still no
theory which would predict robustly the experimentally observed
behavior, especially for canonical flows such as Couette,
Poiseuille and Hagen-Poiseuille flows, where linear spectral
stability analysis is known to fail \cite{Drazin}.

The current experimental evidence suggests that the transition
from laminar to turbulent state has the character of a
finite-amplitude instability, cf. the recent work of Mullin and
Peixinho \cite{Mullin} and figure 7 in that reference. However,
based on the natural limitations of any realistic experimental
procedure, such as inability to introduce disturbances of all
possible forms and to wait infinitely long to see if a particular
disturbance will lead to transition (the pipe is of finite
length!), one has to keep in mind that the actual transition may
turn out to be not a finite-amplitude instability \footnote{A
classical example of a finite-amplitude instability is the
Takens-Bogdanov bifurcation. Maybe we should give both the
equation and its phase portrait?}, but an infinitesimal
instability in a Lyapunov sense \footnote{May be we should give a
precise definition and a visual illustration to contrast it to the
finite-amplitude instability case.}. At the same time, due to the
finite length of a pipe, one cannot say for sure that the
disturbances, which seem to grow, will not decay eventually.

Thus, it would be honest to say that the experimental suggestions
cannot be considered as a firm evidence, and the gap between the
natural deficiencies of experiments and the current theories can
only be bridged by careful sorting out and making precise the
theoretical results and by the subsequent punctilious
interpretation.

In this note we make one step in this direction by analyzing some
of the key assumptions in the current theoretical views. The first
one is the standard consideration of disturbances as defined on
domains which are infinite in the streamwise direction, $-\infty <
x < +\infty$, which is usually justified by the translational
invariance of the base plane-parallel state. As we will show in \S
\ref{section:domain_type_effects}, the usage of a more relevant to
experiments domains setup, $0 \ge x < +\infty$, leads to
unexpected stability results. The second one lies at the
foundation of the transient growth approach, namely the energy
conservation by the nonlinear terms of the NSEs. Just to remind
the reader, the arguments usually made in the context of the
transition to turbulence problem are
\begin{enumerate}
\item Nonlinear terms do not produce energy;
\item Thus, in order to explain transition one has to focus on the
linear terms;
\item The linear terms produce energy only transiently and thus
the transient growth, originating in the non-normality of the
linear operator, is the key to explaining the transition, i.e. the
transition is ``essentially linear'' \cite{Trefethen:I} and
non-normality is the necessary condition for subcritical
transition \cite{Henningson:I,Reddy}.
\end{enumerate}
As we will show in \S \ref{section:nonlinear_terms_effects}, the
energy conservation is not true in the context of the transition
to turbulence problems in general. In the rest of the paper, we
discuss other important issues of the current theories, namely the
interrelation of various stability notions, the effects of
infinite dimensionality, the covariant nature of the transition
phenomena and how non-normality of the linear operators and
finite-amplitude instability fit into this picture.

\section{\label{section:domain_type_effects}Domain type effects}

As mentioned in the introduction, the stability analysis of the
Couette and pipe flow is usually performed on domains unbounded in
both directions, $-\infty < x < +\infty$. If one recalls the way
the experiments on the transition are usually done, i.e. one
introduces disturbances at the inlet location and observes how
they evolve downstream, then it becomes clear that the
semi-infinite domain, $x \in [0,+\infty)$, is more relevant as a
mathematical idealization (in reality the domains are finite, of
course). To illustrate the domain type effect on the stability
results, namely whether the domain is infinite or semi-infinite,
we consider first the Kovasznay flow in \S
\ref{subsection:Kovasznay}, which is treatable analytically and
thus allows us to make the key points in the most clear way, and
then the Couette flow in \S \ref{subsection:Couette}, which we
study both numerically and analytically.

\subsection{\label{subsection:Kovasznay}Kovasznay flow}

\begin{figure}
\centering \epsfig{figure=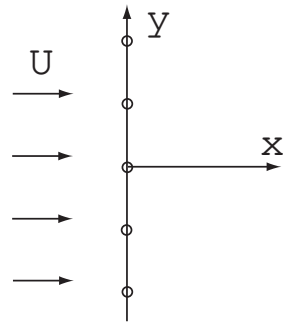,height=1.75in}
\caption{Kovasznay flow.} \label{kovasznay}
\end{figure}

Let us first analyze the inviscid version of the Kovasznay flow,
i.e. the flow behind a periodic grid located at $x=0$, as shown in
figure \ref{kovasznay}. This choice of the flow is due to its
direct relevance to the transition to turbulence problem both in
terms of the plane-parallel base state and the flow domain type,
as well as due to transparent analytical treatment.
In the 2D stream-function formulation, when the velocity
components are $u = \psi_{y}$ and $v = - \psi_{x}$, its dynamics
obeys
\begin{align}
\Delta \psi_{t} + \psi_{y} \Delta \psi_{x} - \psi_{x} \Delta
\psi_{y} = 0,
\end{align}
which after decomposition, $\psi = \Psi + \psi'$, into the basic
state $\Psi(x,y) = y$ periodic for $x>0$ with period $1$, and into
the perturbation $\psi'$ produces the following linearized
evolution equation for $\psi'$:
\begin{align}
\label{equation_linearized_kovasznay} \Delta \psi_{t}^{\prime} +
\Psi_{y} \Delta \psi_{x}^{\prime} = 0.
\end{align}
Performing the standard eigenvalue analysis of
\eqref{equation_linearized_kovasznay}, $\psi'(t,x,y) \rightarrow
\phi(x,y) e^{\lambda t}$, where in view of the solution
periodicity in $y$-direction
\begin{align}
\label{eigenfunction_kovasznay} \phi(x,y) =
\sum_{n=0}^{\infty}{\alpha_{n}(x) \cos{2 \pi n y} + \beta_{n}(x)
\sin{2 \pi n y}},
\end{align}
we find that all the mode amplitudes $\alpha_{n}(x)$ and
$\beta_{n}(x)$ are decoupled, since the operator $\mathcal{L} =
\left(\lambda + \partial_{x}\right) \Delta$ acting on $\phi(x,y)$
is of even order in $\partial_{y}$. Hence,
\begin{align}
\left|\begin{array}{cccc} \mathcal{L}_{0} & \ & \ & 0 \\
\ & \mathcal{L}_{1} & \ & \ \\
\ & \ & \mathcal{L}_{2} & \ \\
0 & \ & \ & \ddots
\end{array}\right| \cdot \left|\begin{array}{c} \alpha_{0} \\
\alpha_{1} \\
\alpha_{2} \\
\vdots
\end{array}\right| = \boldsymbol{0},
\end{align}
that is each ``amplitude'' $\alpha_{n}$ satisfies
\begin{align}
(\lambda + \mathrm{d}_{x}) (\mathrm{d}_{x}^{2} - 4 \pi^{2} n^{2})
\alpha_{n}(x) = 0, \, n \in \Bbb{Z}^{+}.
\end{align}
For example, for $n=0$ let us contrast the solution of this
eigenvalue problem on infinite and semi-infinite domains:
\begin{itemize}
\item $x \in (-\infty,+\infty)$: assuming that $y(x) \in L_{2}$,
by applying Fourier transform we get $\lambda = - i k$, $k \in
\Bbb{R}$, i.e. marginal stability.

\item $x \in [0,+\infty)$ and $|y(x)| < \infty$: clearly,
instability is present since there is an eigenfunction $y \sim
e^{- \mu x}$, $\mu \ge 0$ such that the eigenvalue $\lambda =
\mu$, which clearly leads to instability.
\end{itemize}
Same type of analysis can be done in the viscous case as well.

\begin{figure}
\centering \epsfig{figure=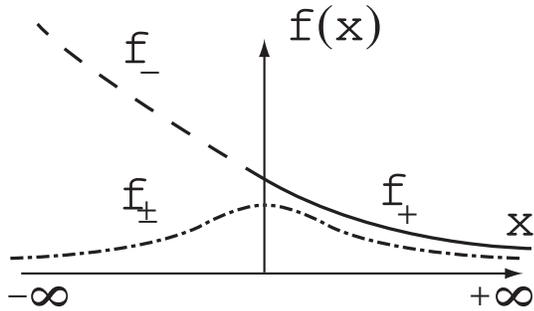,height=1.75in}
\caption{Semi- versus infinite flow domains. $f_{+}$ is defined on
$x \in [0,+\infty)$, $f_{-}$ is defined on $x \in (-\infty,0]$,
$f$ is defined on $x \in (-\infty,+\infty)$}
\label{semi_vs_infinite}
\end{figure}

This counter-intuitive difference in the stability results between
the semi-infinite and infinite domains can be appreciated with the
sketch in figure \ref{semi_vs_infinite}. Namely, if, for example,
one restricts (eigen-) functions $f(x)$ to be bounded for all $x$
including infinities as motivated by the physics, then the space
of functions defined on $x \in (-\infty,+\infty)$ is more
restricted compared to the space of functions defined on $x \in
[0,+\infty)$. Indeed, if one can construct a function $f_{+}$
bounded on $x \in [0,+\infty)$ which also satisfies the
Orr-Sommerfeld (OS) equation
\eqref{Orr_Sommerfeld_Couette_equation}, then continuation of this
function onto $x \in (-\infty,0]$ may lead to an unbounded
function $f_{-}$, as dictated by the structure of the linear
Orr-Sommerfeld operator and as illustrated in figure
\ref{semi_vs_infinite}.
Note that this explains the sensitivity of the critical
bifurcation (Reynolds) number to the properties of disturbances at
the domain inlet, $x=0$: while their amplitudes do not play a role
in view of the linearity of the problem, gradient-like properties
of the disturbances do! Indeed, varying these properties of
disturbances at $x=0$, which may be masked by their amplitude,
effectively changes the boundary conditions at $x=0$ and thus the
size of the function space. Since restricting the domain to $x \in
[0,+\infty)$ enlarges the function space, one can expect that the
spectrum enlarges as well and may lead to instabilities.

Finally, just to reiterate on the crucial distinction of the
stability on infinite versus semi-infinite domain, we remind the
reader that an instability on a semi-infinite domain implies a
growth of some eigenmode, say function $f_{+}$ defined only on a
semi-infinite domain in figure \ref{semi_vs_infinite}. Then, if
the domain is infinite, the unstable eigenmode $f_{+}$ obtained
for a semi-infinite domain is not defined in general (cannot be
continued for negative $x$ in the original function space, say the
space of bounded function) and cannot grow in the part of the
domain, where it is not defined.

\subsection{\label{subsection:Couette}Couette flow}

One can expect that the analogous behavior should take place in
the Couette and other ``troublesome'' flows. First recall that in
the case of the Couette flow, Romanov \cite{Romanov} rigorously
proved that the disturbance defined on $x \in (-\infty,+\infty)$
and decaying at $x = \pm \infty$ can not lead to instability.

Let us show that the Couette flow is unstable on a semi-infinite
domain as opposed to the case considered on an infinite domain. In
the latter case, when the disturbance is defined on $-\infty < x <
+\infty$ and decays at infinities, it is known \cite{Romanov} that
the upper bound of the real parts of the spectrum of the linear
operator $A$ is $\sup \mathrm{Re}(\lambda) \le - c / Re$ with $c >
0$. If, on the other hand, analogously to the Kovasznay flow we
consider the eigenfunctions $\phi(x,y)$ defined on $x \in
[0,\infty)$, then assuming the separated form $\phi(x,y) = a(y)
b(x) = a(y) e^{- \mu x}$ with $\mu \ge 0$ \footnote{Alternatively,
one can apply a cosine transform, which would be a more systematic
way of dealing with the problem, but our goal here is just to
demonstrate the presence of instability, but not to explore the
instability picture exhaustively.}, we get the following
eigenvalue problem for $a(y)$:
\begin{align}
\lambda a_{yy} = Re^{-1} a_{yyyy}, \nonumber \\
y=-1,1: \, a = a_{y} = 0, \nonumber
\end{align}
which is given in the case $\mu = 0$ (which according to the
Fourier analysis should be present if the disturbance does not
decay at $x = \infty$). The straightforward dispersion relation
for this equation yields
\begin{align}
\lambda = 4 \pi^{2} n^{2} \, Re^{-1}, \, n \in \Bbb{N}, \nonumber
\end{align}
i.e. there is the eigenvalue $\lambda = 0$ which gives marginal
stability for any $Re$, as opposed to the result on the infinite
domain.

While the above is the rigorous result on the stability of the
Couette flow when the disturbances do not decay at infinity, one
can apply heuristic Synge's method to the general case, $\mu \ge
0$:
\begin{subequations}
\label{Orr_Sommerfeld_Couette}
\begin{align}
\label{Orr_Sommerfeld_Couette_equation} \left[{\mathrm{d}^{2}
\over \mathrm{d}y^{2}} + \mu^{2} - Re \left(\lambda - \mu y\right)
\right] \left({\mathrm{d}^{2} \over \mathrm{d}y^{2}} +
\mu^{2}\right) a(y) = 0, \\
\label{Orr_Sommerfeld_Couette_BCs} y=-1,1: \, a = a_{y} = 0.
\end{align}
\end{subequations}
Multiplying by $a^{*}$, integrating over $y \in [-1,1]$, and using
the notation $I_{i}^{2} = \int_{-1}^{1}{|a^{(i)}|^{2} \,
\mathrm{d}y}$, we get
\begin{align}
&Re \, \lambda (\mu^{2} I_{0}^{2} - I_{1}^{2}) = I_{2}^{2} - 2
\mu^{2} I_{1}^{2} + \mu^{4} I_{0}^{2} \nonumber \\
- &Re \, \mu \left[\int_{0}^{1}{y a_{y}^{2} \, \mathrm{d}y} -
\mu^{2} \int_{-1}^{1}{y a^{2} \, \mathrm{d}y}\right], \nonumber
\end{align}
which in the limit $\mu \rightarrow + \infty$, physically
corresponding to the localized disturbance at the inlet, yields
\begin{align}
\label{asymptotics_Synge_like} \lambda \sim Re^{-1} \mu^{2} > 0,
\end{align}
i.e. one should observe an instability. The above asymptotics is,
of course, valid only if the solution $a(y)$ does not have a
strong dependence on $\mu$. While, as we will see shortly, $a(y)$
does depend on $\mu$, \eqref{asymptotics_Synge_like} turns out to
give the right insight.

Rigorous dispersion relation for the Couette flow on semi-infinite
domain can, in fact, be written down analytically. If we rewrite
\eqref{Orr_Sommerfeld_Couette} in the operator form, $L_{1} L_{2}
a(y)$, where $L_{1} = {\mathrm{d}^{2} \over \mathrm{d}y^{2}} +
\mu^{2} - Re \left(\lambda - \mu y\right)$ and $L_{2} =
{\mathrm{d}^{2} \over \mathrm{d}y^{2}} + \mu^{2}$, then we can
exploit this factorization of the original forth-order operator in
deriving the dispersion relation. First note that the kernel of
these operators without boundary conditions are
\begin{align}
\mathrm{ker} \, L_{1} &= \mathrm{span}\left\{\mathrm{Ai}(\ldots),
\mathrm{Bi}(\ldots)\right\}, \nonumber \\
\mathrm{ker} \, L_{2} &= \mathrm{span}\left\{\sin(\ldots),
\cos(\ldots)\right\}. \nonumber
\end{align}
Thus, effectively we are solving
\begin{align}
\label{Orr_Sommerfeld_Couette_reduced} L_{2}^{00} a(y) = \alpha
\mathrm{Ai}(\ldots) + \beta \mathrm{Bi}(\ldots),
\end{align}
where $L_{2}^{00}$ is the smaller operator with the boundary
conditions \eqref{Orr_Sommerfeld_Couette_BCs}. Thus, in order to
solve \eqref{Orr_Sommerfeld_Couette_reduced} one must have
\begin{align}
\alpha \mathrm{Ai}(\ldots) + \beta \mathrm{Bi}(\ldots) \bot
\mathrm{ker} \, L_{2}^{00*}, \nonumber
\end{align}
where $L_{2}^{00*}$ is the adjoint (not formally adjoint!) to
$L_{2}^{00}$. Simple usage of the definition of adjoint
\begin{align}
\langle v, L_{2}^{00} w \rangle = \langle L_{2}^{00*} v, w \rangle
\equiv \langle L_{2} v, w \rangle
\end{align}
shows that $L_{2}^{00*}$ coincides with $L_{2}$, i.e. the operator
without boundary conditions! Since we know the kernel of $L_{2}$,
then for equation \eqref{Orr_Sommerfeld_Couette_reduced} and thus
\eqref{Orr_Sommerfeld_Couette} to have a solution, it is necessary
and sufficient that the Gram determinant vanishes:
\begin{align}
\left|\begin{array}{cc} \left<\mathrm{Ai}|\cos\right> & \left<\mathrm{Ai}|\sin\right> \\
\left<\mathrm{Bi}|\cos\right> & \left<\mathrm{Bi}|\sin\right>
\end{array}\right| = 0. \nonumber
\end{align}
In order to get some useful insights into the structure of a
solution, let us analyze one of the entries of the above
dispersion relation, e.g.
\begin{align}
\left<\mathrm{Ai}|\cos\right> = \int_{-1}^{1}{\mathrm{Ai}(-\alpha
y + \beta) \, \cos{\mu y} \, \mathrm{d}y}, \nonumber
\end{align}
where
\begin{align}
\alpha = (\mu \mathrm{Re})^{1/3}, \, \beta = \left({\lambda \over
\mu} - {\mu \over Re}\right) \alpha. \nonumber
\end{align}
As suggested by the asymptotics \eqref{asymptotics_Synge_like}, it
is tempting to consider the limit $\mu \gg 1$ and thus to exploit
the stationary phase method, i.e. by considering $\cos{\mu y}$ as
fast oscillating function and $\mathrm{Ai}(-\alpha y + \beta)$ as
a slowly varying function. However, it turns out that for the
given range of $y$ the periods of oscillation of both Airy and
cosine functions are of the same order, which invalidates
application of the stationary phase method. Another possible
approach is to consider the Airy function with large arguments
$|-\alpha y + \beta| \gg 1$, but then either $\lambda \sim O(1)$
leads to inconsistent asymptotics or $|\lambda| \gg 1$ invalidates
the assumption of a uniform large argument $|-\alpha y + \beta|
\gg 1$ for all $y$. In any case, large values of $\mu$ indeed turn
out to produce interesting structure of the solution, as
illustrated in figure \ref{eigenfunction_Re5000mu40} for $\mu$;
higher values of $\mu$ increase the number of oscillations in the
``accordion'' structure.

In view of this fundamental difficulty to resolve this problem
with the available rigorous analytical methods, we appealed to the
numerical solution of \eqref{Orr_Sommerfeld_Couette} by expanding
the solution into functions based on the complete set of the
Chebyshev polynomials $T_{n}(y) = \cos{n \mathrm{arccos} \, y}$
\begin{align}
a(y) = \sum_{i=0}^{N}{c_{i} \, \phi_{i}(y)},
\end{align}
which guarantee the convergence faster than any power of $N$
\cite{Orszag}. Here the basis functions are given by
\begin{align}
\phi_{i}(y) = T_{i}(y) - 2 {i+2 \over i+3} T_{i+2}(y) + {i+1 \over
i+3} T_{i+4}(y), \nonumber
\end{align}
i.e. they automatically satisfy the boundary conditions
\eqref{Orr_Sommerfeld_Couette_BCs}. All the inner products
associated with the Galerkin projection can be computed
analytically, which is another advantage of this method.

\begin{figure}[!ht]
\centering \ \hspace{-0.4cm}\subfigure[Spectrum (eigenvalues
continue to the negative part of the real
axis.]{\epsfig{figure=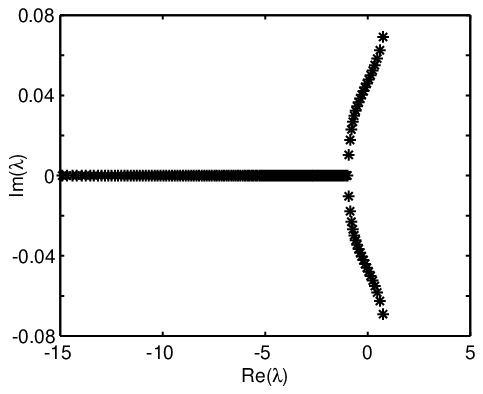,width=2.0in}\label{spectrum_Re5000mu1}}
\qquad \qquad \subfigure[Most unstable
eigenfunction.]{\epsfig{figure=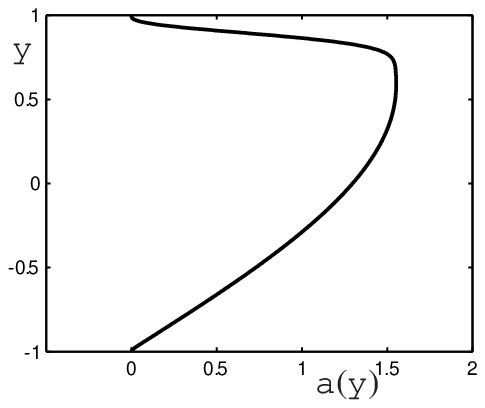,width=2.0in}\label{eigenfunction_Re5000mu1}}
\caption{Spectral picture for the Couette flow at $Re=5000$ and
$\mu=1$. The rightmost eigenvalue has real part $\mathrm{Re}
\lambda_{\mathrm{max}} \simeq 0.75988755$ and the corresponding
eigenfunction is shown in figure \ref{eigenfunction_Re5000mu1}.}
\label{embedding}
\end{figure}
Since the goal here is not to study the complete bifurcation
picture for the Couette flow, but to demonstrate that it is
spectrally unstable for some finite Reynolds number, say $Re=5000$
and $\mu=1$. The resulting spectrum is shown in figure
\ref{spectrum_Re5000mu1} and the eigenfunction corresponding to
the eigenvalue with the largest real part in figure
\ref{eigenfunction_Re5000mu1}.
\begin{figure} \centering
\epsfig{figure=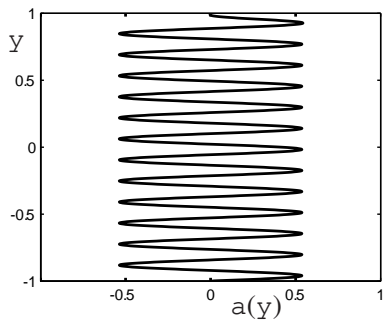,width=2.0in}
\caption{The eigenfunction corresponding to the rightmost
eigenvalue has real part $\mathrm{Re} \, \lambda_{\mathrm{max}}
\simeq 38.5314476$ for the Couette flow at $Re=5000$ and
$\mu=40$.} \label{eigenfunction_Re5000mu40}
\end{figure}
As figure \ref{spectrum_Re5000mu1} clearly suggests, the
transition in the Couette flow is of Hopf bifurcation type. For
each fixed $\mu$ one can compute a critical Reynolds number
$Re_{c}$, e.g. for $\mu = 1$ we get $Re_{c} \simeq 75$. As $\mu$
increases, i.e. the disturbance is localized around the inlet, the
value of $\mathrm{Re} \, \lambda_{\mathrm{max}}$ increases as
well, as anticipated \eqref{asymptotics_Synge_like} and in
agreement with the general observation at the end of \S
\ref{subsection:Kovasznay}. Note that there is nothing wrong with
unbounded growth rates for infinite values of $\mu$, which is a
common feature in many fundamental hydrodynamic instabilities in
the short-wave limit, e.g. Rayleigh-Taylor instability. Of course,
in the nonlinear setting one does not observe an infinite growth
rates, which is suppressed by the nonlinear effects, which can
also be dissipative as will be shown in \S
\ref{section:nonlinear_terms_effects}. Also with increasing $\mu$
the eigenfunction structure becomes more complicated as we
predicted in the above analytical study, e.g. for $Re=5000$ and
$\mu=40$ the eigenfunction is shown in figure
\ref{eigenfunction_Re5000mu40}. More complete study of the
stability picture for the Couette flow is out of the scope of this
work and will be presented elsewhere.

\section{\label{section:nonlinear_terms_effects}Nonlinear terms effects}

As pointed out in the introduction, the usual argument made is
that the linear terms in the NSEs are the only ones which produce
energy and thus the nonlinear terms are energy-conserving and only
mix energy between different modes
\cite{Grossmann:II,Henningson:II,Henningson:I}. The claim made in
\cite{Henningson:II} in the context of discussing shear flows with
base state velocity field $\mathbf{U}$ is based on the
consideration of the evolution of the kinetic energy
$\label{kinetic_energy} E(t) = \|\mathbf{u}\|^{2} / 2$ of the
disturbance velocity field $\mathbf{u}$ either over domains with
periodic (i.e., compact domains) or homogeneous (zero) boundary
conditions for disturbance, which leads to the classical
Reynolds-Orr equation
\begin{align}
\label{equation_RO} {\mathrm{d} E \over \mathrm{d} t} = &- \nu
\|\nabla \mathbf{u}\|^{2} - \left<D_{ij},u_{i} u_{j}\right>,
\end{align}
where $D_{ij}$ is the symmetric part of the base state velocity
gradient tensor $\nabla \mathbf{U}$ \footnote{The usual way of
using this equation to predict stability characteristics is to
maximize the right-hand side for all possible perturbation
solutions (under the constraint of divergence-free fields) using
the calculus of variations \cite{Straughan}. Therefore, the
conclusions of this method are often too conservative, since the
allowed perturbation fields do not necessarily satisfy the NSEs.}.
The linearity of \eqref{equation_RO}, i.e. absence of the
nonlinear terms effects, motivated the ``transient-growth'' idea
of focusing only on the linear mechanisms.

The problem with the above argument is that the ``troublesome''
shear flows are defined not on either over domains with periodic
(i.e., compact domains) or homogeneous (zero) boundary conditions
for disturbance, but on domains with at least one semi- or
unbounded dimension and the boundary condition of boundedness of
the disturbance field at infinity. Thus, let us re-derive equation
\eqref{equation_RO} taking into account that the boundary
conditions on $\partial \Omega$ could be moved away to infinity.
This leads to the following equation \footnote{I'd prefer to
rewrite this equation in a vector not a tensor form.}
\begin{align}
\label{equation_RO_full} &- {\mathrm{d} E \over \mathrm{d} t} =
\int\limits_{\partial \Omega}{n_{i} u_{i} p \, \mathrm{d}s} + \nu
\int\limits_{\partial \Omega}{n_{j} u_{i}
u_{i,j} \, \mathrm{d}s} + \nu \|\nabla \mathbf{u}\|^{2} \\
&\left<D_{ij},u_{i} u_{j}\right> + \boxed{{1 \over 2}
\int_{\partial \Omega}{n_{j} u_{j} u_{i} u_{i} \, \mathrm{d}s}} +
{1 \over 2} \int_{\Omega}{n_{j} U_{j} u_{i} u_{i} \, \mathrm{d}x},
\nonumber
\end{align}
where $\mathbf{n}$ is the normal outward (w.r.t. $\partial
\Omega$) vector and part or the whole of $\partial \Omega$ may be
at infinity. If the domain $\Omega$ is unbounded, as in many
applications, then the evaluation of the boundary terms in
\eqref{equation_RO_full} becomes non-trivial, since it depends on
the rate of the solutions decay in unbounded spatial directions
\footnote{For example, if some function $f(r,\theta)$ is
integrated along the ``boundary'' of a plane, then the
corresponding integral can be represented as $\int_{0}^{2
\pi}{f(r,\theta) \cdot r \, \mathrm{d}\theta}$, which is finite
when $r \rightarrow + \infty$, only if the exponent is the
majorant, $|f(r,\theta)| < M \cdot r^{-\alpha}$, is $\alpha \ge
1$.}, and therefore the nonlinear terms (cubic term in
\eqref{equation_RO_full}) do not disappear, in general. In fact,
there are no reasons to expect that these terms vanish in the
Couette or other channel and pipe flows, since the disturbance, if
it leads to an instability, does not decay as it propagates from
the entrance to infinity. Moreover, if we consider the flow on a
semi-infinite domain, as discussed in \S
\ref{section:domain_type_effects}, then there is a non-zero
contribution of the nonlinear terms at this portion of $\partial
\Omega$ as well. As for open shear flows \footnote{In the case of
the Couette and pipe flows it is obvious that the unbounded
$x$-direction is homogeneous and the all the eigenmodes have the
structure $\sim e^{i k x}, \ k \in \Bbb{R}$, that is bounded at
infinities. There is no justification to decompose the strip-like
domain into periodic boxes; after all, when we solve the heat
equation on a line we never do that and should we do this
procedure, we would fail to construct the solution. While the
solution of the heat equation on the real line obviously decays,
the existence of periodic solutions of the full NSEs
\cite{Nagata:I} in the case of the Couette flow proves the
possibility of non-localized solutions of the NSEs in these
situations.} the nonlinear terms do not vanish since it is
well-known that in order to construct a full set of eigenfunctions
from the Orr-Sommerfeld equation, it is necessary to relax the
homogeneous boundary conditions at infinity to the condition of
boundedness at infinity, $|\mathbf{u}|_{|x| \rightarrow \infty} <
\infty$, \cite{Salwen,Grosch,Herron:I}, because the homogeneous
boundary condition leads only to a finite set of a point discrete
spectrum and therefore the inclusion of the continuous spectrum
component is necessary for completeness. The completeness is of
course important if one wants to study the question of stability
rigorously. Therefore, one has to admit that the role of the
nonlinear terms in the NSEs for shear flow problems is not a
simple mixing of energy, but the disturbance energy can be
denerated/dissipated by the nonlinearities.

Finally, another very important point is that the usual assertion
that at early times \textit{the nonlinear effects are not
important} is not necessarily true since, for example, the linear
term $\left<D_{ij},u_{i} u_{j}\right>$ and nonlinear one $- {1
\over 2} \int_{\partial \Omega}{n_{j} u_{j} u_{i} u_{i} \,
\mathrm{d}s}$ in equation \eqref{equation_RO_full} may become
comparable even at small times either thanks to the very small
values of $D_{ij}$ and/or the large cumulative effect of nonlinear
terms when integrated over $\partial \Omega$ (possibly infinitely
extended). In any case, no one has ever proved that this cannot
occur for the flows in question!

It should be kept in mind that the energy is a nonlocal measure of
the fluid motion, while in reality we are interested in the
pointwise description, since we do not know much about the
singular structure of the NSEs solution.

\section{Other issues}

For the purpose of the subsequent discussion discussion, it is
convenient to treat the NSEs \eqref{NSEs} as an
infinite-dimensional ODE in the operator form in some Banach space
$X$:
\begin{align}
\label{main_equation_nonlinear} {\mathrm{d} \mathbf{u} \over
\mathrm{d} t} = A \mathbf{u} + N(\mathbf{u}),
\end{align}
where $A$ is the linear operator, usually stationary, and $N$ is
the nonlinear operator. Historically, this approach allowed one to
apply the dynamical systems methods of Dalekii \& Krein
\cite{Krein}, Yudovich \cite{Yudovich:I} for nonlinear stability
of the equilibrium solutions of \eqref{main_equation_nonlinear}.
Equation \eqref{main_equation_nonlinear} can be obtained from
\eqref{NSEs} by utilizing the Helmholtz decomposition
\cite{Chorin}, i.e. through decomposing the solution vector field
into the sum of its divergence- and curl-free components, e.g.
$L^{p}(\Omega)=S^{p}(\Omega) \otimes G^{p}(\Omega)$, where
$S^{p}(\Omega)$ is the Banach space obtained by closing the set of
solenoidal vectors and $G^{p}(\Omega)$ is the Banach space
obtained by closing the set of gradients in the norm of
$L^{p}(\Omega)$. Then one can introduce the Leray projector
$\Bbb{P}$, which projects any vector in $L^{p}(\Omega)$ onto
$S^{p}(\Omega)$, and thus yields \eqref{main_equation_nonlinear}.
In particular, if the stability of a nontrivial stationary
(time-independent) base state $\mathbf{U}(\mathbf{x})$ is studied,
then the linear and nonlinear operators in
\eqref{main_equation_nonlinear} take the form
\begin{subequations}
\begin{align}
\label{operator_linearized_NSEs} A \mathbf{u} &= \Bbb{P} \left[-
\mathbf{U} \cdot \nabla \mathbf{u} - \mathbf{u} \cdot \nabla
\mathbf{U} + Re^{-1} \Delta u\right], \\
\label{operator_nonlinear_NSEs} N(\mathbf{u}) &= - \Bbb{P}
\left[\nabla \cdot (\mathbf{u} \otimes \mathbf{u})\right].
\end{align}
\end{subequations}
As follows from \eqref{operator_linearized_NSEs}, one can see the
linear NSEs operator $A$ as a non-self-adjoint perturbation of the
Laplace operator.

\subsection{\label{subsection:stability_notions}Interrelation of stability notions}

The notions of stability or instability, as physically observable
phenomena, were given by Lyapunov \cite{Khalil}. Namely, the
equilibrium solution (base state) $\mathbf{u}^{e} \in X$ of
\eqref{main_equation_nonlinear} is said to be \textit{Lyapunov
stable} (sometimes called nonlinearly stable since
\eqref{main_equation_nonlinear} is nonlinear) if for any $\epsilon
> 0$ there exists a $\delta
> 0$ so that the initial conditions $\mathbf{u}_{0} \in X$ and
$\|\mathbf{u}_{0} - \mathbf{u}^{e}(t_{0})\|_{X} < \delta$ imply
that (i) there exists a solution
$\mathbf{u}(t,t_{0},\mathbf{u}_{0})$ and (ii)
$\|\mathbf{u}(t,t_{0},\mathbf{u}_{0}) - \mathbf{u}^{e}(t)\|_{X} <
\epsilon$ for all $t$. Then, the notion of \textit{Lyapunov
instability} is simply the negation of the above definition of
stability. While this negation formally does not require existence
to hold, i.e. the solution seizing to exist is a particular form
of instability, from the physical point of view one does need
existence in $X$ to get a sensible instability result
\cite{Krechetnikov:II}. Note that the \textit{linear stability} is
the Lyapunov stability of the linearized version of
\eqref{main_equation_nonlinear}, $\mathbf{u}_{t} = A \mathbf{u}$
in the sense of the above definition. This should also be
distinguished from the \textit{spectral stability}, i.e. a formal
notion obtained from the spectral problem, $A \mathbf{v} = \lambda
\mathbf{v}$, associated with the linearized version of
\eqref{main_equation_nonlinear}, when the spectrum is in the left
half-plane.

There is a common view that linear eigenvalue analysis implies
nonlinear Lyapunov stability, e.g. \cite{Henningson:I} who, in the
context of their discussion of the transition to turbulence, refer
to Sattinger \cite{Sattinger} in order to claim that a connection
between linear and nonlinear stability has been established.
However, in that work Sattinger demonstrated stability using the
Reynolds-Orr equation, that is in a non-pointwise norm, for
compact domains only, while the shear flows are usually considered
on domains with at least one extended dimension. In general, as
one can gather from most of the fluid dynamics literature, it is
used as \textit{a rule}: if the spectrum of the linearized
operator \eqref{operator_linearized_NSEs} is in the left half
plane then one has stability and the instability takes place if
there are eigenvalues in the right half plane. Below we first
discuss the interrelation of all these notions in finite
dimensions, and then address infinite dimensions in \S
\ref{subsection:infinite_dim_effects}.

\begin{Example}
First, consider a conservative system with the Hamiltonian $H(p,q)
= {p^{2} / 2} + V(q)$, where the potential is quartic $V(q) =
q^{4} / 4$. As one can immediately see from the Hamilton's
equations, $q_{t}=p$ and $p_{t} = -q^{3}$, the linear and
nonlinear stability of its equilibrium, $q = p = 0$, definitions
do not imply each other: this Hamiltonian system demonstrates
nonlinear stability because the energy function is concave, but
its linearization around the origin, $\dot{p} = 0$ \& $\dot{q} =
p$, produces a solution growing linearly in time, i.e. it is
linearly unstable. However, this example is spectrally stable;
thus, spectral stability does not imply even linear stability,
however the converse is true.
\end{Example}

One might think that the above is a peculiar feature of the
Hamiltonian dynamics, but in fact one observes the same for
dissipative systems, e.g. for $q_{t}=p$ and $p_{t} = -q^{3} -
p^{3}$.

Another example, discussed in \cite{Pollard} (see also
\cite{Siegel}) proves that linear stability does not imply
nonlinear stability:

\begin{Example} Given the Hamiltonian
\begin{align}
H = {1 \over 2} (p_{1}^{2} + q_{1}^{2}) - (p_{2}^{2} + q_{2}^{2})
+ {1 \over 2} p_{2} (p_{1}^{2} - q_{1}^{2}) - q_{1} q_{2} p_{1}.
\nonumber
\end{align}
from the corresponding Hamilton's equations, we find that the
origin, $\mathbf{q}=\mathbf{0}$ and $\mathbf{p}=\mathbf{0}$, is
linearly stable, but a one-parameter, $\tau$, family of solutions
for this system
\begin{align}
p_{1} &= \sqrt{2} \, {\sin(t-\tau) \over t-\tau}, & \ p_{2} &=
{\sin{2(t-\tau)} \over t-\tau}, \nonumber \\
q_{1} &= -\sqrt{2} \, {\cos{(t-\tau) \over t-\tau}}, & \ q_{2} &=
{\cos{2(t-\tau)} \over t-\tau}, \nonumber
\end{align}
demonstrates nonlinear instability by blowing up at some finite
time, $t=\tau$.
\end{Example}

Therefore, from the above examples we conclude that
\begin{itemize}
\item linear instability does not imply nonlinear instability;

\item linear stability does not imply nonlinear stability;

\item spectral stability does not imply both linear and nonlinear
stability,
\end{itemize}
However, spectral instability implies nonlinear instability in
finite dimensions based on the classical Lyapunov indirect method
\cite{Khalil}.

\subsection{\label{subsection:infinite_dim_effects}Infinite dimensionality effects}

In infinite dimensions, the situation is even richer, because the
latter property -- the Lyapunov indirect method -- does not hold
in general. As mentioned in \S \ref{subsection:stability_notions},
the usual approach is to base the conclusion of stability or
instability of \eqref{main_equation_nonlinear} on the spectrum
$\sigma(A)$ of the linear operator $A$: if all eigenvalues lie
strictly (i.e. no eigenvalues are on the imaginary axis) in the
left half of the complex plane $\Bbb{C}$, then the zero solution
is stable, while if there are eigenvalues in the left half-plane
of $\Bbb{C}$, then the zero solution is unstable.

Let us discuss the question of linearized stability systematically
\footnote{I think it would be nice to connect the discussion of
semi-groups with non-normality in the infinite-dimensional case.}.
The solution of the linearized version of
\eqref{main_equation_nonlinear} which can be written symbolically
as
\begin{align}
\label{solution_linear} \mathbf{u}(t) = \mathbf{u}(t_{0}) \ e^{A
(t-t_{0})}.
\end{align}
For a Banach space $X$ of finite dimension it is well known that
if $\sup\limits_{\mathrm{Re} \, \lambda}{\lambda \in \sigma(A)}$
is $ < 0$, then $\|T(t)\|$ decays exponentially \cite{Pazy}. This
behavior is a consequence of the fact that linear operators in
finite-dimensional Banach spaces have only point spectrum. Since
this is not the case in infinite-dimensional Banach spaces one
does not expect this result to be true in general. From a formal
point of view, the exponent $e^{A t}$ makes mathematical sense if
the operator $A$ is bounded, i.e. $\|A x\|_{X} \le M \|x\|_{X}$
(in future we drop the index denoting the space $X$ in which the
norm applies unless it becomes important to distinguish), so that
\begin{align}
\label{exponent_series_representation} e^{A t} =
\sum\limits_{k=0}^{\infty}{A^{k} t^{k} / k!},
\end{align}
is well-defined, and from \textit{the spectral mapping theorem}
\cite{Engel} one can connect the spectra $\sigma(A)$ and
$\sigma(e^{A t})$ via
\begin{align}
\label{spectrum_bounded_operator} \sigma(e^{A t})=e^{t \sigma(A)}.
\end{align}
The convergence of \eqref{exponent_series_representation} allows
one to get a rough estimate, $\|e^{A t}\| \le
\sum\limits_{k=0}^{\infty}{\|A\|^{k} t^{k} / k!}=e^{\|A\| t}$,
$t>0$. However, the knowledge of the spectrum $\sigma(A)$ allows
one to get sharper estimates: in particular, if $\|e^{A}\| \le q$,
the spectrum $\sigma(A)$ lies in the interior of $\mathrm{Re} \,
\lambda \le \log{q}$. If, however, the operator $A$ is unbounded,
which is usually the case in hydrodynamics, then $e^{A t}$ makes
only symbolic sense and the equality is not true in general
\cite{Phillips} (see also Lax \cite{Lax:I}, pp. 434--439) and, in
fact, the spectral mapping property
\eqref{spectrum_bounded_operator} does not hold, i.e.
\begin{align}
\label{spectrum_unbounded_operator} \sigma(e^{A t}) \supset e^{t
\sigma(A)},
\end{align}
from where it follows that one cannot conclude on stability of
\eqref{main_equation_nonlinear} based on the knowledge of spectrum
of $\sigma(A)$. For unbounded operator the property
\eqref{spectrum_bounded_operator} takes place only if it is an
infinitesimal generator of an \textit{analytic semi-group}, the
convenient test for which is via proving that the operator is
\textit{sectorial} (\textit{cf}. Henry \cite{Henry}), i.e. if its
eigenvalues are contained in a cone sector with an apex angle $<
\pi$ in the left-half of the complex plane.

These known facts can be most conveniently formulated with the
help of two notions: \textit{spectral bound}, $r(A) =
\sup\left\{\mathrm{Re} \, \lambda | \, \lambda \in
\sigma(A)\right\}$, and \textit{the growth abscissa} of the
semigroup, $\omega(A) = \lim\limits_{t \rightarrow \infty}{t^{-1}
\log\left(\|e^{A t}\|\right)}$. Then, in finite dimensions it is
true that $\omega(A)=r(A)$, which is implicit in all studies of
hydrodynamic stability. In the infinite-dimensional case, this
equality is not true in general, as will be discussed below.
However, some operators in certain spaces do possess this
property, such as the Stokes operator is known \cite{Yudovich:I}
to generate an analytic semigroup in any $L^{p}$ space. Thus, one
can expect that if \eqref{operator_linearized_NSEs} is not a
strong enough perturbation of the elliptic operator, which also
depends on the boundary conditions, then the operator should still
be a generator of an analytic semigroup.

On the other hand, there are many stability problems where the
diffusive mechanisms, as controlled by elliptic operators (the
Stokes, biharmonic, etc.), may be negligible, and thus $\omega(A)
\neq r(A)$. For example, hyperbolic PDEs exhibit such behavior, as
can be illustrated with the following simple example due to
Renardy \cite{Renardy:I}

\begin{Example} Consider the following
hyperbolic equation:
\begin{align}
\label{example_Renardy} u_{tt} = u_{xx} + u_{yy} + e^{i y} u_{x},
\end{align}
with $2 \pi$-periodic boundary conditions in both directions.
Introducing $v = u_{t}$, the semigroup operator $A$ associated
with \eqref{example_Renardy} is $A(u,v) = \left(v,u_{xx} + u_{yy}
+ e^{i y} u_{x}\right)$, which in fact generates a
$C_{0}$-semigroup in the space $H^{1} \times L^{2}$ for the pair
$(u,u_{t})$. The analysis \cite{Renardy:I} demonstrates that $r(A)
= 0$, while $\omega(A) = 1/2$. \label{example_Renardy_txt}
\end{Example}
In general, it is well-known after the work of Zabczyk
\cite{Zabczyk} that for any two real numbers $\omega_{s} <
\omega_{0}$ there exists a strongly continuous semigroup
$\{T_{t}\}_{t \ge 0}$ on a Hilbert space, such that $\omega_{s} =
\sup\left\{\mathrm{Re} \, \lambda; \, \lambda \in
\sigma(A)\right\}$ and $\left|T_{t}\right| = e^{\omega_{0} t}, \ t
\ge 0$, where $\sigma(A)$ denotes the spectrum of the generator
$A$ of the semigroup $\{T_{t}\}_{t \ge 0}$.

The above most systematically can be understood in terms of
classification of semi-groups \cite{Pazy}, namely
\textit{parabolic} versus \textit{hyperbolic}, as resulted from
the application of the abstract semi-group theory to evolution
partial differential equations. In the parabolic case the
operators are the infinitesimal generators of an analytic
semi-group, while in the hyperbolic case one can expect the
behavior as in the above example \ref{example_Renardy_txt} due to
Renardy \cite{Renardy:I}. While behavior of example
\ref{example_Renardy_txt} is common for Hamiltonian systems, in
which the eigenvalues are distributed symmetrically w.r.t both
axes in the complex plane and the placement of all eigenvalues on
the imaginary axis is a necessary but not a sufficient condition
for stability of an equilibrium, equation \eqref{example_Renardy}
is clearly non-conservative, which suggests that such pathologies
in the linear stability picture are common for dissipative systems
too.

First of all, the existence and (in)stability of the solutions of
the full nonlinear system \eqref{main_equation_nonlinear} are most
conveniently (at the current level of mathematics) studied by
transforming it into the integral form via Duhamel's formula
\begin{align}
\label{Duhamel_formula} \mathbf{u}(t) = \mathcal{T}(\mathbf{u}) =
\mathbf{u}(0) e^{A t} + \int\limits_{0}^{t}{e^{A (t-\tau)} \,
N(\mathbf{u}(\tau)) \, \mathrm{d}\tau},
\end{align}
which is not only used to solve \eqref{main_equation_nonlinear}
using fixed point theorems for the mapping
$\mathcal{T}(\mathbf{u})$, but also clearly illustrates the
importance of both the linear terms -- the semigroup $\{e^{A
t}\}_{t \ge 0}$ -- and the nonlinear terms in infinite dimensions.
Indeed, on any relevant to this discussion Banach space $X$ the
nonlinear terms will be unbounded in view of the presence of
unbounded operator of differentiation in
\eqref{operator_nonlinear_NSEs}, which can be compensated only by
the smoothing effect of the semigroup $e^{A (t-\tau)}$. The latter
can happen, of course, only if the semigroup is ``nice'' enough.
This illustrates the importance of consideration of both linear
and nonlinear effects when studying the questions of stability.
Formula \eqref{Duhamel_formula} is the basis of all rigorous
stability studies, such as due to Krein \cite{Krein}, Henry
\cite{Henry}, and Yudovich \cite{Yudovich:I}.

Thus, it is not surprising that infinite dimensionality may lead
to pathologies, which are not possible in finite dimensions. For
example, in the context of elasticity problems it is known that
the positive definite second variation of the energy (and thus the
corresponding system should be nonlinearly stable according to the
Dirichlet theorem, valid in finite dimensions only) does not
guarantee the stability in view of the possible presence of an
infinite number of unstable directions \cite{Ball:I}.

Another complication arising due to infinite dimensionality is the
norm-dependence of stability criteria -- the issue which has been
understood for a long time \cite{Yudovich:I,Friedlander:II}. In
the finite-dimensional case this difficulty cannot arise since all
norms in finite-dimensional Banach spaces are equivalent.

\begin{Example}
One of the simplest examples of this subtlety is due to
\cite{Yudovich:I} and represents a linear PDE:
\begin{align}
{\partial u \over \partial t} = x {\partial u \over \partial x},
\nonumber \\
u(0,x)=\phi(x), \nonumber
\end{align}
the unique solution of which is $u(x,t)=\phi\left(x \
e^{t}\right)$. Thus one can express the $L^{p}$-norm of the
solution derivatives via $\left\|{\partial^{k} u(\cdot,t) \over
\partial x^{k}}\right\|_{L^{p}(\Bbb{R})} = e^{(k - p^{-1}) t}
\left\|\phi^{(k)}\right\|_{L^{p}(\Bbb{R})}$, and therefore one has
(a) asymptotic stability in $L^{p}(\Bbb{R})$ for $1 \le p <
\infty$, (b) Lyapunov stability in $L^{\infty}(\Bbb{R})$, and (c)
exponential instability in Sobolev spaces $W^{k,p}(\Bbb{R})$ with
$k>1$, $p \ge 1$ or $k=1$, $p>1$.
\end{Example}

Finally, quite often by ``nonlinear stability'' one understands
``energy stability'', which is not quite correct, since nonlinear
stability is a stability in the Lyapunov pointwise sense, while
the energy norm is a global measure \footnote{Maybe it would help
if we give a nice pathological example?}. In order to make a
pointwise sense of the energy-like norms, so that the function
belongs to the H\"{o}lder space of $k$-times continuously
differentiable functions $f \in C^{k}$, one needs energy-like
bounds for the solution derivatives, that is $f$ should belong to
the Hilbert space $H^{k}$ with high enough index, so that one can
employ the Sobolev inequality; e.g. in the two-space dimensions:
\begin{align}
\left|f\right|_{C^{k}} \le C \|f\|_{H^{k+s}}, \ s \ge 1. \nonumber
\end{align}
From here it follows that establishing bounds on the usual kinetic
energy norm $H^{0}$ is not enough to assert the bounds on the
solution, and thus to claim the stability of the solutions in the
Lyapunov sense.

\subsection{Non-normality and covariance}

Starting with Trefethen et al. \cite{Trefethen:I} the following
two-dimensional model is very popular \cite{Henningson:I} for
illustrating what is ``supposed'' to happen in the transition
problem
\begin{align}
\label{model_Trefethen} {\mathrm{d} \mathbf{u} \over \mathrm{d} t}
= A \mathbf{u} + \mathbf{F}(\mathbf{u}), \ A =
\left(\begin{array}{cc} - a_{1} & 1 \\ 0 & - a_{2}
\end{array}\right),
\end{align}
where $A$ is a non-normal operator in a sense that $A \, A^{*}
\neq A^{*} \, A$, where $A^{*}$ is the adjoint operator, with
$a_{1,2} \sim Re^{-1}$ being small exponents determining the
time-evolution, and $\mathbf{F}(\mathbf{u})$ is a nonlinear energy
conserving operator and. Since the linear operator $A$ is
non-normal, then its eigenvectors corresponding to the eigenvalues
$\lambda_{1}=- a_{1}$ and $\lambda_{2}=- a_{2}$ are almost
parallel for $Re \gg 1$, which in the case used by Trefethen et
al. \cite{Trefethen:I}, $a_{1}=Re^{-1}$ and $a_{2}=2 \, Re^{-1}$,
are:
\begin{align}
\label{model_Trefethen_eigenvectors} \mathbf{e}_{1} =
\left[\begin{array}{c} 1
\\ 0\end{array}\right],
\ \mathbf{e}_{2} = \left[\begin{array}{c} 1 \\
- Re^{-1}\end{array}\right].
\end{align}
In the limit $Re \rightarrow \infty$ the matrix $A$ becomes the
non-trivial Jordan block, when the solution $\mathbf{u}$ grows
algebraically. Due to this non-normality for finite $Re$, the
system $\dot{\mathbf{u}} = A \mathbf{u}$ experiences a transient
growth for $t = O(Re)$, e.g. for the initial conditions
$\mathbf{u}(0) = \left[0, 1\right]^{T}$:
\begin{align}
\mathbf{u}(t) &= \left[\begin{array}{c} Re \\ 0\end{array}\right]
e^{- {t \over Re}} + \left[\begin{array}{c} -Re \\
1\end{array}\right]
e^{- {2 t \over Re}} \nonumber \\
\label{example_transient_growth}
&= \left[\begin{array}{c} t + O\left(t^{2} \over Re\right) \\
1 - {2 t \over Re} + O\left(t^{2} \over
Re^{2}\right)\end{array}\right].
\end{align}
Next, the key dynamic feature of all these models is the presence
of a finite-amplitude instability; in the case of Trefethen et al.
\cite{Trefethen:I}
the nonlinearity is $\|\mathbf{u}\| B \mathbf{u}$ with $B = \left(\begin{array}{cc} 0 & -1 \\
1 & 0\end{array}\right)$, which has three stable nodes (one at the
origin) and two saddle points, while Manneville
\cite{Manneville:I} used $\mathbf{F}(\mathbf{u}) = \left(u v, -
u^{2}\right)^{T}$, which has two stable nodes (one is at the
origin) and a saddle. Manneville used example
\eqref{model_Trefethen} to advocate that the non-normality of the
linear operator is not in itself responsible for the by-pass
transition, while Trefethen et al. \cite{Trefethen:I} insisted
that the transition is ``essentially linear''. While the form of
Trefethen's nonlinearity was criticized \cite{Waleffe:I} as not
relevant to the NSEs, it is clear that it cannot be obtained as a
result of some kind of reduction from the NSEs, since the square
root is not Taylor expansion-friendly.

If one expects that transition is a fundamental physical
phenomenon, then it must be accounted in a covariant
(coordinate-free) manner, that is its understanding should be
independent of the coordinate system. Thus, let see if model
\eqref{model_Trefethen} possesses this property.

In this subsection we consider the linear part of
\eqref{model_Trefethen} recalling some elementary facts, while in
the next section we discuss the fully nonlinear dynamics. Even
though the original matrix $A$ in \eqref{model_Trefethen} is
non-normal, there exists a transformation which makes this linear
operator in \eqref{model_Trefethen} normal. Clearly, normal
matrices are a subset of the diagonizable matrices and the
relation between normal and diagonizable matrices is
well-understood \cite{Mitchell}. The notion of
\textit{diagonizability} is an intrinsic notion (that is,
independent of a coordinate system) as opposed to to
\textit{non-normality}, since the latter notion depends upon a
system of coordinates. In terms of genericity notion, \textit{cf}.
Wiggins \cite{Wiggins}, i.e., informally, how common are the
physical systems with non-normal operators, it is obvious that
non-normal matrices form a dense subset in the set $C^{n^{2}}$ of
all $n \times n$ matrices. Further, we will comment on the
infinite-dimensional operators, but first consider the finite
dimensional case as relevant to the discussion of model
\eqref{model_Trefethen}. In fact, this clear understanding of
which notions are intrinsic or not shades some light on the claims
that the non-normal linear operators is the key to transition.
Appealing to the standard matrix theory \cite{Franklin}, let us
consider a linear transformation $A$ on the Euclidean space
$\Bbb{E}^{n}$, which maps the vector $\mathbf{x}$ into
$\mathbf{x}'$: $\mathbf{x}' = A \mathbf{x}$. If matrix $T$
consists of the basis vectors
$\mathbf{t}^{1},\ldots,\mathbf{t}^{n}$, then $\mathbf{x} = T
\mathbf{y}$ and $\mathbf{x}' = T \mathbf{y}'$, where $\mathbf{y}$
and $\mathbf{y}'$ are components of the corresponding vectors
$\mathbf{x}$ and $\mathbf{x}'$. Note that $\mathrm{det} \, T \neq
0$ since it forms a basis. Then, the relation between $\mathbf{y}$
and $\mathbf{y}'$ is simply $\mathbf{y}' = T^{-1} \, A \, T \,
\mathbf{y}$. Matrices $A$ and $T^{-1} \, A \, T$ are similar, and
$T^{-1} \, A \, T$ can be made diagonal (and thus normal) if and
only if $A$ has $n$ linearly independent eigenvectors \footnote{It
should be kept in mind that the reduction of matrix to its Jordan
normal form is an unstable operation \cite{Arnold:I}.} (in
particular, if $A$ has distinct eigenvalues, but not necessarily).
Of course, if the matrix $A$ is originally normal, then it will
have orthogonal eigenvectors and thus $T$ will be orthogonal,
which means that normality will be `preserved'. Anyway, non-normal
matrices may be reduced to normal ones under appropriate change of
coordinates. Since we looking for a covariant description of
physical phenomena, i.e. independence of our understanding of the
phenomena of a coordinate system, then it becomes clear that
non-normality has nothing to do with the covariant understanding
of the fundamental cause of the transition. While the transient
growth effects may be important exactly as ``transient'' effects,
which can be most clearly seen through the singular value
decomposition, on the time scales greater than the transient
growth time, e.g. $t > Re^{-1}$ in
\eqref{example_transient_growth}, their effect is not relevant in
any coordinate system. Becides these general remarks, one should
also point out from the positions of control theory that not all
non-normal operators can lead to significant transient effects,
but only those which possess some high sensitivity subspaces in
their domain of definition together with the presence of
disturbances in this subspace.

The above discussion for operators in finite-dimensional spaces
can be translated to the case of infinite-dimensional operators,
as we discuss here in the context of PDEs. Clearly, the linear
operator in the NSEs \eqref{operator_linearized_NSEs} is
non-normal for non-trivial base states $\mathbf{U}(\mathbf{x})$.
The degree of non-normality of \eqref{operator_linearized_NSEs}
increases as the Reynolds number increases since the self-adjoint
part of \eqref{operator_linearized_NSEs} \footnote{The Laplacian
is a self-adjoint operator under special requirements on the
boundary conditions, e.g. homogeneous ones.}, that is the
Laplacian, becomes less dominant. Most clearly, this probably can
be seen in a spectral space after the Galerkin projection. In any
case, the non-normality of the partial differential operator leads
to a transient in time growth in the same way as a non-normal
matrix-operator, which is a self-evident fact, but nevertheless
quite a number of works were devoted to illustrate this behavior
in the context of the NSEs \cite{Reddy, Bamieh}.

\subsection{Finite-amplitude effects}

From the nonlinear covariant analysis viewpoint, it is tempting to
perform the normal form analysis of \eqref{model_Trefethen} in
order to reveal the universal behavior of \eqref{model_Trefethen}.
However, the local nature of the normal form analysis, i.e.
inability to get the global bifurcation picture, does not allow
one to capture the finite-amplitude instabilities inherent in
\eqref{model_Trefethen}: e.g. in the non-resonant case of $a_{1}$
and $a_{2}$ in \eqref{model_Trefethen}, the normal form is simply
\begin{align}
{\mathrm{d} \mathbf{x} \over \mathrm{d} t} = J \mathbf{x}, \ J =
\left(\begin{array}{cc} - a_{1} & 0
\\ 0 & - a_{2}
\end{array}\right), \nonumber
\end{align}
and in the resonant case, $a_{2} = 2 a_{1}$ is
\begin{align}
{\mathrm{d} \mathbf{x} \over \mathrm{d} t} = J \mathbf{x} +
\left(\begin{array}{c} 0 \\ a x_{1}^{2}
\end{array}\right), \, a \in \Bbb{R}. \nonumber
\end{align}
Both the above normal forms do not exhibit finite-amplitude
instabilities. This fundamental inability of the normal form
analysis to capture the global bifurcation picture can be
understood with the following simple example, which also leads to
some further insights.

\begin{Example}
Let us consider the following equation
\begin{align}
\label{original_equation} {\mathrm{d} x \over \mathrm{d} t} =
-\epsilon x + x^{2},
\end{align}
which describes transcritical bifurcation and exhibits a
finite-amplitude instability with the critical amplitude $x_{c} =
\epsilon$. With the goal to reduce this equation to a linear
equation
\begin{align}
\label{normal_form} {\mathrm{d} y \over \mathrm{d} t} = -\epsilon
y,
\end{align}
let us introduce a transformation, $x = y + F(y)$, where $F(y)$ is
a function to be determined. Substitution of this transformation
into \eqref{original_equation} gives
\begin{align}
\label{original_equation_transformed} {\mathrm{d} y \over
\mathrm{d} t} [1 + F'(y)] = -\epsilon [y + F(y)] + [y + F(y)]^{2}.
\end{align}
In the standard treatment of the local normal form analysis (e.g.
\cite{Wiggins}), one assumes that $|F'(y)| \ll 1$ and thus inverts
$[1 + F'(y)]$ approximately, i.e. $[1 + F'(y)]^{-1} \simeq 1 -
F'(y)$. In fact, one does not need this approximate procedure in
order to get an equation for $F(y)$: instead, one multiplies
\eqref{normal_form} by $F'(y)$ and subtracts from
\eqref{original_equation_transformed}:
\begin{align}
{\mathrm{d} y \over \mathrm{d} t} = -\epsilon y - \epsilon F(y) +
\epsilon y F'(y) + [y + F(y)]^{2},
\end{align}
that is, $F(y)$ is determined from
\begin{align}
\epsilon [F - y F'] = [y + F]^{2},
\end{align}
with the result
\begin{align}
F(y) = {\epsilon y \over y - \alpha} - y,
\end{align}
where $\alpha \neq 0$ for the transformation $x = y + F(y)$ to be
non-singular. Thus, for $x \neq \epsilon$ the transformation $x =
y + F(y)$ ``straightens out'' the trajectories of the original
system \eqref{original_equation} and produces a trivial dynamics
as in \eqref{normal_form}, but is singular for $x=\epsilon$.
\end{Example}
The latter property, i.e. singularity of the transformation which
linearizes the dynamics globally except for at the threshold
amplitudes can be used to identify the presence and location of
the finite-amplitude instabilities in the appropriate
configuration (phase) spaces.

The indicated in the introduction possibility of finite-amplitude
instability nature of the transition suggested a search for
finite-amplitude solutions in the shear flows. For example, Nagata
\cite{Nagata:I} found a finite-magnitude periodic solutions in the
Couette flow, which coexists with the linear Couette profile,
appealing to the concept \textit{bifurcation from infinity}
\cite{Rosenblat}. The travelling wave-like solutions are known
both in the context of the pipe \cite{Faisst} and plane Couette
\cite{Nagata:II} flows. However, their ``strongly unstable
character'', since they are all saddle points in phase space
\cite{Kerswell:I}, is also well-known \cite{Eckhardt}. The latter
work also advocates that the turbulent state in pipe flow
corresponds to a chaotic saddle (unstable aperiodic orbit) in
state space. The idea is that travelling wave solutions presumably
constitute a `skeleton' about which complicated time-dependent
orbits may drape themselves temporarily before falling back to the
laminar state \cite{Kerswell:I}.

Waleffe \cite{Waleffe:I} advocated the idea of exploring the size
of the domain of attraction of the finite-amplitude solutions
following the original thoughts of Orr and Thomson. Also,
Trefethen et al. \cite{Trefethen:I} conjectured that the
non-normality of the linear operator shrinks the size of the
attraction basin and, in fact, the threshold amplitude scales as
$Re^{-\gamma}$.

The above dynamical systems approach is based on the
finite-dimensional view of the NSEs dynamics. The latter is
usually justified \cite{Kerswell:I} by the argument that the
motion of a viscous fluid in a finite domain is always
finite-dimensional which is due to the viscous cutoff of fine
scales with reference to \cite{Constantin}. However, the
troublesome flows always have at least one unbounded dimension,
which undermines the above logic and poses the question on the
possible crucial effect of infinite-dimensionality. Next, as
currently understood, all these finite-amplitude solutions proved
to be unstable. Moreover, it is very likely that if the transition
is indeed a finite-amplitude instability, the attracting solutions
in fact are intrinsically time-dependent or even chaotic and
occupy some subset of the phase space. Thus, there is little hope
to find the corresponding solutions analytically.

\section{Discussion}

The purpose of this note was to bring a number of important
theoretical issues to the attention of the fluids community,
solving of which may help to make some progress on a reasonable
qualitative understanding and quantitative prediction of the
experimental observations, which are still lacking. In particular,
as follows from \S \ref{section:domain_type_effects} and \S
\ref{section:nonlinear_terms_effects}, exploring the effects of
the domain type and of the energy conservation by nonlinear terms
are the first natural issues to address systematically.

The authors also see two other possible ways of exploring the
problem of transition rigorously. The first way is to develop a
semi-local nonlinear stability theory, which could be a natural
generalization of the stability theory due to Krein, Yudovich, and
Henry by inclusion of the finite-amplitude instability picture in
consideration. The second possible avenue is to gain insights into
the geometric structure of the equations, which allow one to
identify the possible regions of attraction in the phase space and
thus to establish the stability picture in large (i.e. both local,
global, finite-amplitude, etc.). Namely, one first explores the
phase space structure of the Hamiltonian approximation (ideal
fluid) and, if this is not sufficient for explaining the
instability picture, one adds dissipative effects (viscosity). In
particular, this would allow one to understand if the fundamental
basis of the transition phenomena is Hamiltonian or intrinsically
due to dissipative (viscous) effects; see also
\cite{Krechetnikov:II}. The reader can easily appreciate the
effectiveness of this approach with the help of, say, the
Takens-Bogdanov system \cite{Lewis}:
\begin{align}
\ddot{x} = \mu - x^{2} = - V_{x}, \ V(x) = -\mu x + {x^{3} / 3}.
\end{align}
Namely, locating maxima and minima of the potential $V(x)$ (or
alternatively of the Hamiltonian) indicates the stability picture.
This idea, of course, is very old and goes back to Lagrange,
Dirichlet and other classics. Next, one adds dissipative effects
\cite{Krechetnikov} and observes how the stability picture
changes. The infinite-dimensional case is, of course, more
complicated, but some progress has been done in this direction as
well \cite{Krechetnikov:II}.

Finally, while all the previous fluid mechanics experience over
the last few centuries suggests  that the NSEs are the adequate
description of fluid motion, strictly speaking one cannot discard
the chance that the NSEs do not describe the subtle transition to
turbulence phenomena. One can name many reasons for which the NSEs
may turn out to be inappropriate for modeling of the transition.
For example, since the transition is a phenomenon presumably very
sensitive to initial and boundary conditions, then it should be
very sensitive to the details of the equations. Since Newtonian
fluid description is an approximate one, then in the conventional
NSEs for incompressible fluid we discard by hyperbolic effects
common for non-Newtonian fluids. Same can be said about the
incompressibility approximation. While one might argue that these
effects may have an influence only at some marginal time and
spatial scales, the history of hydrodynamics knows a number of
fundamental examples when small effects affect the flow in the
large (Prandtl's boundary layer theory, for instance).

\section{Acknowledgements}

R.K. would like to thank the attendees of the session on
transition in shear flows at the Second Canada-France Congress
(June 1-6, 2008), where the presented here results were announced,
for the feedback.

\end{document}